\documentclass[twocolumn,prl,aps]{revtex4}
\usepackage{graphicx}

\newcommand{\be}{\begin{equation}}
\newcommand{\ee}{\end{equation}}
\newcommand{\bea}{\begin{eqnarray}}
\newcommand{\eea}{\end{eqnarray}}
\newcommand{\HH}{{\cal H}}

\newcommand{\la}{\langle}
\newcommand{\ra}{\rangle}
\newcommand{\lb}{\left[}
\newcommand{\rb}{\right]}
\newcommand{\lp}{\left(}
\newcommand{\rp}{\right)}
\renewcommand{\vec}[1]{{\bf #1}}

\begin{document}

\title{The Higgs resonance in fermionic pairing}

\author{Roman Barankov} 

\affiliation{Department of Physics, Boston University, Boston, Massachusetts 02215, USA}
\begin{abstract} 

The Higgs boson in fermionic condensates with the BCS pairing interaction
describes the dynamics of the pairing amplitude. I show that the existence and
properties of this mode are sensitive to the energy dispersion of the
interaction. Specifically, when the pairing is suppressed at the Fermi level,
the Higgs mode may become unphysical (virtual) state or a resonance with
finite lifetime, depending on the details of interaction. Conversely, the
Higgs mode is discrete for the pairing interaction enhanced at the Fermi
level. This work illustrates conceptual difficulties associated with
introducing collective variables in the many-body pairing dynamics.

\end{abstract} \pacs{}
\keywords{}

\maketitle

The low-energy properties of many-body systems with spontaneously broken
symmetries are encoded in the collective modes associated with the structure
of the order parameter. The dynamics is usually dominated by the gapless
Goldstone modes that restore the symmetry~\cite{Goldstone62}, which makes them
easily observable in variety of systems. In contrast, the massive Higgs
modes~\cite{Higgs64} resulting from the amplitude variations of the order
parameter are hard to detect. Moreover, they do not necessarily exist when the
order parameter is generated dynamically due to collective effects.
Understanding how these modes emerge in a microscopic framework is essential
for constructing field-theoretical descriptions operating with a partial set
of dynamical variables.

This peculiarity of the Higgs mode becomes evident in the BCS model of
superconductivity~\cite{BCS57}, where pairing of fermions with opposite spins
and momenta is described by a complex-valued order parameter. Several
collective modes distinct from the Higgs mode have been considered in
superconductors: the excitonic states~\cite{Vaks61,Bardasis61}, the
high-temperature phase modes~\cite{Carlson75}, and also a phonon gap mode
observed in materials with coexisting superconductivity and charge-density
waves~\cite{Sooryakumar80,Littlewood81}. At the same time, no direct
signatures of the intrinsic amplitude mode have been found so far. The main
difficulty has to do with the structure of quasi-particle excitations in the
system. 

The role of quasi-particle states is clarified in the response to a small
pairing perturbation. The induced dynamics can be described as the dephasing
of individual quasi-particle states excited in the vicinity of the Fermi
energy, with the rate of dephasing proportional to the excitation energy. As a
result one finds a power-law decay of the perturbation~\cite{Volkov74}.
Equivalently, this behavior can be linked to the square-root singularity of
the density of states at the gap energy~\cite{Volkov74}. Recent extension of
the analysis to non-linear regimes~\cite{Barankov06-2,Yuzbashyan06} has also
demonstrated the absence of the intrinsic collective excitations of the
pairing amplitude. One of the implications of this behavior is the lack of an
effective description of the pairing dynamics such as time-dependent
Ginzburg-Landau theory in gapped superconductors~\cite{Abrahams66,Gorkov68}.
However, the results of the analysis rely on the key assumption of constant
interaction between fermions and change qualitatively otherwise.

\begin{figure}[t]
\includegraphics[width=3.5in]{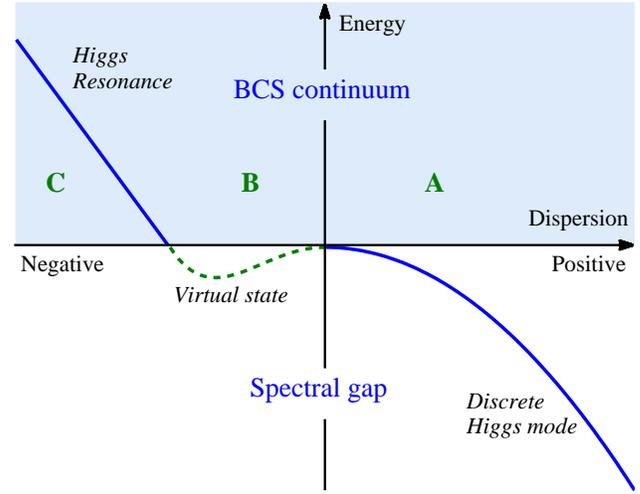}
\vspace{-0.5cm} 
\caption[]{
The Higgs mode energy sketched as a function of dispersion: the mode is
discrete within the spectral gap for positive dispersion (regime A); the mode
is non-physical (virtual state) for a smooth negative dispersion (regime B);
and it becomes a resonance inside the continuum of the BCS quasi-particles for
sharp and negative dispersion (regime C).
}
\vspace{-0.5cm}
\label{fig:Higgs_diagram}
\end{figure}

In this work, I discuss three different regimes of the pairing dynamics that
appear due to the energy dispersion of fermionic interaction in the BCS model.
The main results of this work are summarized in Fig.~\ref{fig:Higgs_diagram}.
In particular, the energy of the discrete amplitude mode lies inside the
spectral gap (regime A) resulting in non-decaying dynamics when the pairing is
enhanced at the Fermi energy (positive dispersion). Upon smooth reduction of
the dispersion, the energy of the mode gradually approaches the continuum of
quasi-particle excitations eventually merging with the edge of the spectrum at
constant interaction. In this case, and also for a smooth suppression of the
pairing (negative dispersion) the collective mode becomes non-physical
(virtual state), which is reflected in a power-law decay of the dynamics
(regime B). The mode becomes a resonance (damped oscillation) within the
quasi-particle continuum with a finite lifetime only when the energy-scale of
suppression is small compared to the equilibrium gap at the Fermi energy
(regime C). The lifetime of the Higgs mode is small compared to the
quasi-particle relaxation time, making the dynamics analogous to the Landau
damping~\cite{Landau46} in collisionless plasma.

Our system is described by the BCS Hamiltonian
\be\label{eq:Ham}
\HH=\sum_{\vec p\sigma}\epsilon_\vec p a^\dagger_{\vec p\sigma}a_{\vec
p\sigma}-\sum_{\vec p\vec q}\lambda_{\vec p\vec q}a^\dagger_{\vec
p\uparrow}a^\dagger_{-\vec p\downarrow}a_{\vec q\downarrow}a_{-\vec
q\uparrow},
\ee
where $a^\dagger_{\vec p\sigma}$ is the creation operator of a spin-$1/2$
fermion with momentum state $\vec p$ and spin state
$\sigma=\uparrow,\downarrow$, $\epsilon_\vec p=p^2/(2m)-\mu$ is the energy of
free fermions with respect to the Fermi energy $\mu$. The summation is
conventionally performed over the states within the band $|\epsilon_\vec
p|<\omega_D$ of finite width related to e.g. the Debye energy. 

The energy dispersion of interaction $\lambda_{\vec p\vec q}$ in
Eq.(\ref{eq:Ham}) leads to the energy-dependence of the pairing gap. The
analytical treatment is simplified in the case when it is taken in a
factorized form $\lambda_{\vec p\vec q}=\lambda f_\vec p f_\vec q$, such that
$f_{|\vec p|=p_F}=1$. Diagonalizing the BCS Hamiltonian~(\ref{eq:Ham}), one
obtains the quasi-particle excitation spectrum $E_\vec p=(\epsilon^2_\vec
p+\Delta^2_\vec p)^{1/2}$ characterized by the pairing amplitude
\be
\Delta_\vec p=f_\vec p\Delta_F,
\ee
where the energy dependence is provided by $f_\vec p$, and $\Delta_F$ is the
equilibrium value of the pairing gap at the Fermi energy. The self-consistency
condition takes on the form $2/\lambda_0=\sum_\vec p f^2_\vec p/E_\vec p$ and
defines $\Delta_F$ as a function of $f_\vec p$ and the coupling constant
$\lambda_0$. The factorized form of energy-dependent interaction simplifies
the analysis. Nonetheless, the qualitative results obtained in this work also
apply to a generic form of interaction.

We analyze the dynamics induced by a small perturbation
of pairing interaction, $\delta\lambda(t)=\lambda(t)-\lambda_0$. It is
characterized by the expectation value of the pairing operator
\be
\hat\Delta=\lambda_0\sum_\vec p f_\vec p a^\dag_{\vec
p\uparrow} a^\dag_{-\vec p\downarrow},
\ee
calculated with respect to time-dependent many-body state. The problem is solved by employing the time-dependent mean-field
approach~\cite{Barankov04-1} valid at time scales smaller than the
quasi-particle relaxation time. The linear response of the pairing amplitude
$\delta\Delta_\vec p=f_\vec p\delta\Delta$ is given by
\be\label{eq:linear_response}
\delta\Delta(t)=-i\int_{-\infty}^{t}d\tau \la [\hat\Delta(t),\hat
V(\tau)]\ra,
\ee
where the operators evolve according to the unperturbed
Hamiltonian~(\ref{eq:Ham}), and the average $\la...\ra$ is taken with respect
to its ground state. The perturbation operator
\be\label{eq:perturbation}
\hat V(t)=-\frac{\Delta_F}{\lambda_0}\lp
\frac{\delta\lambda(t)}{\lambda_0}+\frac{\delta\Delta(t)}{\Delta_F}\rp\hat
\Delta^\dag(t)+{\rm
h.c.}
\ee
includes a term $\propto\delta\Delta(t)$ obtained from the averaging of the
interaction energy within the mean-field approximation.

The solution of Eq.~(\ref{eq:linear_response}) simplifies if one assumes the
particle-hole symmetry. This condition implies that the energy-dependent
pairing gap is symmetric with respect to the Fermi energy. In this case, the
phase and the amplitude modes of $\Delta$ decouple and can be analyzed
separately. Concentrating on the amplitude mode $\delta\Delta'$ excited by the
perturbation~(\ref{eq:perturbation}), we obtain the solution of the
inhomogeneous integral equation by the Fourier transformation:
\be\label{eq:Delta_omega}
\delta\Delta'(\omega)=\lb G(\omega)-1\rb\frac{\Delta_F}{\lambda_0}
\delta\lambda(\omega),
\ee 
where $G(\omega)$ is a retarded spectral function describing the many-body
response. By the relation $G^{-1}(\omega)=1+K(\omega)$ it is expressed through
a correlation function $K(\omega)$ that reads in real time as
\be
iK(t)=\frac{\theta(t)}{\lambda_0}\lp\la[\hat\Delta(t),\hat\Delta^\dag(0)]\ra+
\la[\hat\Delta(t),\hat\Delta(0)]\ra\rp.
\ee
Calculating $K(t)$ we derive the expression for the spectral function
\be\label{eq:G_A}
G^{-1}(\omega)=g\int_0^{\omega_D} d\epsilon\,
\lp\frac{\Delta_\epsilon}{\Delta_F}\rp^2
\frac{\omega^2/4-\Delta^2_\epsilon}{E_\epsilon(E_\epsilon^2-\omega^2/4)},
\ee
where $g=\nu\lambda_0$ is dimensionless coupling, and the particle-hole
symmetry is enforced by energy-independent density of states $\nu$. The states
in (\ref{eq:G_A}) are labeled by energy $\epsilon$ instead of momentum $\vec
p$, and according to the usual rule $\omega\to\omega+i0$ for the retarded
function. It is worth noting that Eq.~(\ref{eq:G_A}) is in agreement with
Ref.~\cite{Volkov74}.

The analytic structure of $G(\omega)$ in the complex $\omega$-plane defines
the time-dependence of the pairing amplitude $\delta\Delta(t)$. In particular,
we are interested in locating the poles $\omega_{_\Delta}=\omega_0-i\Gamma$ of
$G(\omega)\propto (\omega-\omega_{_\Delta})^{-1}$, or equivalently the simple
roots of $G^{-1}(\omega)$ on the real axis or in the lower-half plane. In the
dynamics the former correspond to non-decaying oscillations with frequency
$\omega_0$, while the latter are characterized by finite lifetime
$\tau=1/\Gamma$.

\begin{figure}[t]
\includegraphics[width=3.1in]{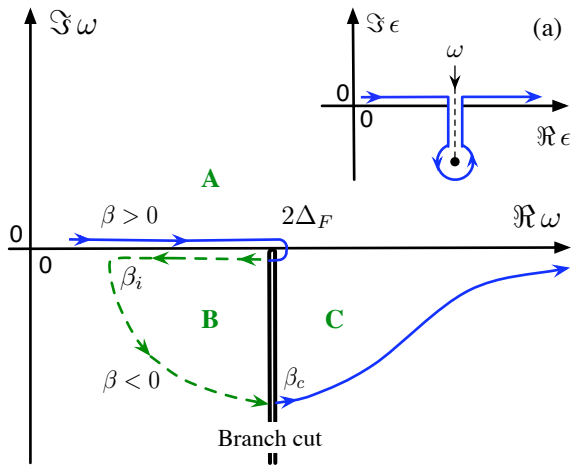}
\vspace{-0.25cm} 
\caption[]{
The analytic structure of $G(\omega)$ sketched in the complex plane of
$\omega$: as $\beta$ changes from positive values to zero, the Higgs pole
approaches the branch point $2\Delta_F$ (regime A); at negative values it goes
into the non-physical sheet (regime B) through the branch cut, and first moves
along the real axis, then obtains imaginary part at $\beta<\beta_i$; at value
$\beta<\beta_c$ it enters the physical sheet as a resonance (regime C). {\it
Inset} (a): Deformation of the integration contour in Eq.~(\ref{eq:G_A})
with $\omega$ crossing the real axis, that results in the analytical structure
shown in the main plot.}
\vspace{-0.25cm}
\label{fig:Higgs_pole}
\end{figure}

Let us first consider the case when the pairing gap 
\be
\Delta_\epsilon^2\approx
\Delta_F^2-\beta \epsilon^2
\ee
has a smooth maximum at the Fermi energy, i.e. the curvature is small
$0\le\beta\ll 1$, and the characteristic energy scale $\gamma$ of the maximum
is large, i.e. $\gamma\gg\Delta_F$. Although in a realistic situation these
two parameters can be related to one another, for the sake of the argument we
consider $\gamma$ and $\beta$ as independent parameters. In this case the root
of $G^{-1}(\omega)=0$ is close to the quasi-particle continuum
$\omega_{_\Delta}\lesssim 2\Delta_F$. By expanding Eq.~(\ref{eq:G_A}) in small
parameter $z=1-\omega^2/(2\Delta_F)^2$, we obtain
\be\label{eq:eigen_1}
G^{-1}(z)\approx -\frac{g\pi}{2}\lp\sqrt{z}-\frac{2}{\pi}\beta \ln\lb
2\gamma/\Delta_F\rb\rp.
\ee
In the dispersionless case $\beta=0$, the spectral function reduces to
$G^{-1}(z)\approx -(g\pi/2) \sqrt{z}$ at $z\gtrsim 0$, and we recover the
square-root singularity that leads to a power-law dephasing discussed in
Ref.~\cite{Volkov74}. When $\beta > 0$, a simple pole of $G(\omega)$ appears
on the real axis of $\omega$ at
\bea\label{eq:Higgs_smooth1}
\omega_{_\Delta}^2\approx 4\Delta_F^2\lp 1-\frac{4}{\pi^2}\beta^2\ln^2
\lb 2\gamma/\Delta_F\rb\rp
\eea
inside the spectral gap of quasi-particle excitations. It is identified as the
discrete Higgs mode in regime A of Fig.~\ref{fig:Higgs_diagram}. Close to the
pole the spectral function
\be\label{eq:G_smooth}
G(\omega)\approx \frac{8}{\pi^2g}\,\beta
\ln\lb 2\gamma/\Delta_F\rb\,\frac{(2\Delta_F)^2}{\omega^2-\omega_{_\Delta}^2}
\ee
defines the spectral weight of the Higgs mode in the dynamics of
$\delta\Delta(t)$. As expected, it vanishes in the dispersionless limit
$\beta=0$. The pairing dynamics at $\beta>0$ is characterized by non-decaying
oscillations of $\delta\Delta(t)$ with frequency~(\ref{eq:Higgs_smooth1}).

By inspecting Eq.~(\ref{eq:Higgs_smooth1}) we find that a virtual state is
formed at smooth negative dispersion. The analytical continuation to negative
$\beta<0$ shows that the mode seemingly has the same energy as for $\beta>0$.
In fact it belongs to a non-physical sheet of $\omega$ (regime B in
Fig.~\ref{fig:Higgs_diagram}) as it follows from careful analysis of
Eq.~(\ref{eq:eigen_1}). Indeed the decrease of $\beta$ from positive values to
zero, brings the energy of the mode to the branch point at $\omega=2\Delta_F$.
Upon further reduction of $\beta$ to negative values, the mode enters the
non-physical sheet of the Riemann surface parameterized by $\omega$ from the
branch cut and moves away from this point along the real axis as a virtual
state (see Fig.~\ref{fig:Higgs_pole}). In the pairing dynamics, the virtual
state leads to a power-law decay of $\delta\Delta(t)$ similar to the
dispersionless case of Ref.~\cite{Volkov74}.

In the opposite limit of sharp dispersion, $\gamma\ll \Delta_F$ and
$|\beta|\gg 1$, this virtual state becomes a resonance within the
quasi-particle spectrum with finite lifetime,
$\omega_{_\Delta}=\omega_0-i\Gamma$. To demonstrate this effect we study a
specific example of the energy-dependent pairing gap:
\be\label{eq:Gap}
\Delta_\epsilon^2=
\Delta_F^2-\beta\frac{\epsilon^2}{1+\epsilon^2/\gamma^2},\quad
f_\epsilon=\Delta_\epsilon/\Delta_F,
\ee
where $\beta$ is a curvature at the Fermi energy, and $\gamma$ is a
characteristic energy scale. We assume $|\beta|\gamma^2\ll \Delta_F^2$, so
that the asymptotic value of the pairing gap
$\Delta_*=\Delta_F(1-\beta\gamma^2/\Delta_F^2)^{1/2}$ is close to $\Delta_F$.
In the case of smooth dispersion, $0<\beta\ll 1$, we obtain
Eqs.~(\ref{eq:Higgs_smooth1}) and (\ref{eq:G_smooth}). In the opposite limit
of sharp energy dispersion $\beta\gg 1$, the energy of the Higgs mode is close
to the asymptotic value of the spectral gap $2\Delta_*$, as it follows from
the expansion of Eq.~(\ref{eq:G_A}):
\be
\omega_{_{\Delta}}^2\approx 4\Delta_*^2+4\,\frac{\int_0^\infty d\epsilon\, 
\frac{\Delta_\epsilon^2-\Delta_*^2}{\epsilon^2+\Delta_\epsilon^2-\Delta_*^2}}
{\int_0^\infty d\epsilon\, \frac{1}{\epsilon^2+\Delta_\epsilon^2-\Delta_*^2}}.
\ee
In this expression, the upper limits of the integrals are substituted with
infinity, by using their fast convergence. Explicit calculation for
model~(\ref{eq:Gap}) reveals a square-root singularity in the energy of the
Higgs mode
\be\label{eq:Higgs_sharp_pos}
\omega_{_\Delta}^2\approx 4\Delta_*^2+4\gamma^2\sqrt{\beta},
\ee
as a function of $\beta$. This singularity is related to the form of
interaction in Eq.(\ref{eq:Gap}), and is not universal.

The spectral weight of the Higgs mode in the dynamics is extracted from the
residue of the pole in the spectral function
\be\label{eq:sharp_weight}
G(\omega)\approx \frac{2}{\pi
g}\,\beta^{1/4}(\gamma/\Delta_F)\frac{(2\Delta_F)^2}{\omega^2-
\omega_{_\Delta}^2},
\ee 
vanishing in the dispersionless limit $\gamma=0$ (see Eq.~(\ref{eq:Gap})). 

The analytic structure of Eq.~(\ref{eq:Higgs_sharp_pos}) is different from
Eq.(\ref{eq:Higgs_smooth1}): as the sign of $\beta$ changes from positive to
negative, the former acquires an imaginary part with the real part
simultaneously pushed in the quasi-particle continuum, since
$\Delta_*>\Delta_F$. To verify this qualitative observation one needs to
continue $G(\omega)$ analytically in the lower-half of complex $\omega$-plane
and locate the pole.

The analytical continuation of $G(\omega)$ is achieved by isolating the
singularities of the integrand in Eq.~(\ref{eq:G_A}). We notice that wherever
the imaginary part of frequency $\Im\,\omega$ changes from positive to
negative value, provided its real part satisfies $\Re\,\omega>2\Delta_F$, a
singularity occurs in the integrand of $G^{-1}(\omega)$ at the energy
$\epsilon$ that solves the equation $E_\epsilon=\Re\,\omega/2$. By deforming
the contour of integration over the energy to enclose the pole, as
Fig.~\ref{fig:Higgs_pole}(a) illustrates, we find the analytical continuation.

Next we apply this method to calculate the energy of the Higgs resonance in
the case of negative dispersion at $|\beta|\gg 1$. The explicit calculation of
$G(\omega)$ for model interaction~(\ref{eq:Gap}) is straightforward but
tedious. Solving for the zeros of the analytic function, we define the energy
of the Higgs resonance
\be\label{eq:Higgs_sharp_neg}
\omega_{_\Delta}^2\approx 4\Delta_*^2-i4\gamma^2\sqrt{|\beta|}.
\ee
As we have anticipated, the energy of the mode coincides with
Eq.~(\ref{eq:Higgs_sharp_pos}) analytically continued to negative values of
$\beta$. The spectral function is obtained from Eq.~(\ref{eq:sharp_weight}) by
substituting $\beta^{1/4}$ with $|\beta|^{1/4}e^{i\pi/4}$.

The analytical structure of $G(\omega)$ for the energy
dispersion~(\ref{eq:Gap}) is shown schematically in Fig.~\ref{fig:Higgs_pole}.
At smooth and negative dispersion below a critical value,
$|\beta|\le|\beta_c(\gamma)|\lesssim 1$, the Higgs mode belongs to the
non-physical sheet (regime B). The critical point $\beta_c(\gamma)$ depends
weakly on the energy scale $\gamma\ll\Delta_F$. In this parameter range there
is an additional critical point $|\beta_i(\gamma)|<|\beta_c(\gamma)|$ such
that at $|\beta|<|\beta_i|$ the energy of the mode is real. Conversely at
$|\beta|>|\beta_i|$ the energy develops imaginary part increasing until the
mode enters the physical sheet at $\beta=\beta_c$ where it becomes a resonance
(regime C).

The change from analytic to non-analytic dependence on $\beta$ in the Higgs
mode energy~(\ref{eq:Higgs_smooth1}) and (\ref{eq:Higgs_sharp_pos}) is
accompanied by a singular behavior of the quasi-particle dephasing. Indeed, by
expanding the quasi-particle energy at small $\epsilon$ we find
$E_\epsilon\approx \Delta_F+(1-\beta)\epsilon^2/(2\Delta_F)$. At $\beta=1$ the
quadratic term in $\epsilon$ vanishes, and the dispersion is controlled by
higher-order terms in the expansion. This singularity affects the
quasi-particle dephasing in the pairing amplitude
$\delta\Delta_{qp}(t)\simeq\Re\int d\epsilon u_\epsilon e^{-i2E_\epsilon t}$,
where $u_\epsilon$ is a smooth function finite at the Fermi energy. Upon
substitution of the expansion of $E_\epsilon$ to the phase factor, we obtain a
$1/2$ power-law decay $\delta\Delta_{qp}(t)\simeq \Re e^{-i2\Delta_Ft}
[(1-\beta)t]^{-1/2}$ valid at long times $t\gtrsim 1/\Delta_F$. Function
$\delta\Delta_{qp}$ has a square-root singularity with a branch cut along the
positive real axis at $\beta\ge 1$. At smaller values the function is
analytic. There is no direct correspondence between this non-analyticity
controlled by the low-energy expansion and the non-analyticity of the Higgs
mode (\ref{eq:Higgs_sharp_pos}) defined by the high-energy asymptotics of the
interaction. The explicit calculation in Eqs.~(\ref{eq:Higgs_sharp_pos}) and
(\ref{eq:Higgs_sharp_neg}) illustrates that the character of the latter is
specific to the model.

The dynamics of the pairing amplitude $\delta\Delta(t)$ is controlled by
$G(\omega)$. The two contributions, from the Higgs pole
$\omega_{_\Delta}=\omega_0-i\Gamma$ located on the physical sheet, and from
the quasi-particle branch cut, compete with each other. At relatively small
times $t\gtrsim 1/\Delta_F$ the quasi-particle states provide the main
contribution leading to the algebraic decay of the pairing amplitude. At
intermediate time scales, $1/\Delta_F\ll t\ll 1/\Gamma$, the Higgs mode
dominates the dynamics characterized by the pairing amplitude oscillating with
frequency $\omega_0$ and decaying exponentially with the decrement
$\tau=1/\Gamma$ (infinite for a discrete Higgs mode). Eventually, at even
longer times $t\gg \tau$ the quasi-particle contribution prevails, although
having by that time a smaller amplitude. When the Higgs mode turns into a
virtual state, the power of algebraic decay is increased compared to the power
$1/2$ discussed above.

The analysis of the pairing dynamics within the BCS model~(\ref{eq:Ham}) is
restricted to time scales small compared to quasi-particle relaxation time
$\tau_{qp}$~\cite{Aronov_review}. At weak coupling and at small temperatures
it is exponentially suppressed, which validates our result for the lifetime of
the Higgs mode~(\ref{eq:Higgs_sharp_neg}) provided $\gamma^2\gg
\Delta_F/(\tau_{qp}\sqrt{|\beta|})$.

In conclusion, the energy dispersion of pairing interaction qualitatively
changes collective dynamics of fermions. The suppression of the interaction at
the Fermi level generically leads to formation of a virtual Higgs state with
energy on a non-physical sheet of Riemann surface. On the other hand, the
enhancement of interaction at the Fermi level makes this mode discrete with
energy located in the spectral gap of quasi-particle excitations. Depending on
the details of interaction, the mode may also become a resonance with finite
lifetime within the quasi-particle continuum.

I am indebted to L.~Levitov, L.~Glazman, and M.~Randeria for insightful discussions of this work.

\end{document}